\documentclass{article}
\usepackage{amsmath}
\usepackage{graphicx} 
\usepackage{tikz-cd}
\usepackage{geometry}
\usetikzlibrary{positioning, arrows.meta}
\usepackage[
    backend=biber,
    style=phys,
    articletitle=false,
    biblabel=brackets
]{biblatex}
\title{Rethinking Collapse: Coupling Quantum States to Classical Bits with quasi-probabilities}

\author{
Dagomir Kaszlikowski\thanks{Corresponding author: phykd@nus.edu.sg}
\and Pawel Kurzynski\thanks{pawel.kurzynski@amu.edu.pl}
}

\date{March 2025}

\begin{document}

\maketitle
\begin{abstract}
 We propose a formulation of quantum measurement within a modified framework of \textit{frames}, in which a
quantum system—a single qubit—is directly coupled to a classical measurement bit. The qubit is represented
as a positive probability distribution over two classical bits, \( a \) and \( a' \), denoted by \( p(aa') \). The
measurement apparatus is described by a classical bit \( \alpha = \pm 1 \), initialized in the pure distribution
\( p(\alpha) = \frac{1}{2}(1 + \alpha) \). The measurement interaction is modeled by a quasi-bistochastic
process \( S(bb'\beta \mid aa'\alpha) \)—a bistochastic map that may include negative transition
probabilities, while acting on an entirely positive state space. When this process acts on the joint initial
state \( p(aa')p(\alpha) \), it produces a collapsed state \( p(bb'\mid\beta) \), yielding the measurement outcome
\( \beta \) with the correct quantum-mechanical probability \( p(\beta) \). This approach bypasses the von
Neumann chain of infinite couplings by treating the measurement register classically, while capturing the
nonclassical nature of measurement through the quasi-bistochastic structure of the interaction.
 \end{abstract}

\section{Introduction}

The quantum measurement problem has long resisted a fully satisfactory resolution. In the standard formulation introduced by von Neumann, measurement is modeled as a unitary interaction between a quantum system and an increasingly large hierarchy of measuring devices, each entangled with the next. This recursive structure leads to the well-known von Neumann chain \cite{vonNeumann1932}, in which the collapse of the wavefunction is never explicitly realized, but continually deferred.

Numerous interpretations and frameworks have been proposed to address this issue. The Copenhagen interpretation introduces a classical/quantum boundary and postulates wavefunction collapse as an axiom when a measurement occurs, without offering a dynamical account. Decoherence-based approaches explain the suppression of interference and the appearance of classicality by tracing out environmental degrees of freedom, but they do not explain the realization of a single outcome \cite{pearle1989,caves2007,zurek2009}. Objective collapse models, such as Ghirardi–Rimini–Weber \cite{ghirardi1986} and refined continuous spontaneous localization \cite{pearle1989}, modify the Schrödinger equation by adding stochastic or nonlinear terms that induce spontaneous localization, at the cost of violating unitarity. In contrast, the many-worlds interpretation preserves unitarity but denies collapse entirely, positing that all outcomes occur in branching universes—raising challenges concerning probability and observer identity. Quantum Bayesianism (QBism) \cite{caves2007subjective} and relational quantum mechanics shift the focus from ontic states to observer-relative information updates, reframing measurement as epistemic, though arguably at the cost of explanatory completeness.

In this work, we propose a formulation of quantum measurement that bypasses the von Neumann chain by coupling a quantum system -- a single qubit -- to a classical measurement bit. The qubit is described as a positive probability distribution over two classical bits ($a = \pm 1, a' = \pm 1$), while the measurement apparatus is represented by a classical bit $\alpha$, initialized in the deterministic state $\alpha = +1$, corresponding to $p(\alpha) = \frac{1}{2}(1+\alpha$). The interaction between system and apparatus is modeled by a quasi-probabilistic process $S(bb'\beta|aa'\alpha$), which yields both the correct quantum measurement statistics and a well-defined post-measurement state.  However, the model relies on transition structures that admit negative probabilities, whose physical interpretation remains generally unresolved \cite{Abramsky_2011,AbramskyBrandenburger2014, HalliwellYearsley2012,deBarros2016,Ryu2019, Feynman1987, Kenfack2004}.

In what follows, we highlight several structural features of the model that will be developed in the rest of
the paper. First, the appearance of negative transition weights can be traced to the combination of a
three-dimensional measurement direction with a single classical bit serving as the measurement register;
this geometric constraint prevents a fully positive representation of all projective measurements in this
minimal setting. Second, we obtain a three-parameter family of quasi-bistochastic maps that all reproduce
the correct single-shot collapse behaviour, and we identify a particularly convenient choice whose repeated
application exhibits a simple cyclic structure and maintains positivity in the scenarios we analyse. Third,
the deterministic preparation of the meter bit, $p(\alpha)=\tfrac12(1+\alpha)$, does not arise as a symmetric informationally complete (SIC)
distribution of any quantum state, so the model necessarily employs initial conditions outside the quantum
state space; this makes explicit why quasi-bistochastic dynamics appear when a classical bit interacts
directly with a qubit. Finally, after tracing out the meter, the resulting transformation acting on the qubit
reduces exactly to the SIC representation of a Lüders projection, so the nonclassical features of the model
are located in the transition map rather than in the state space.


\section{Frame Representations of $n$-Qubit States}

In standard quantum mechanics, the state of an \( n \)-qubit system is represented by a \( 2^n \times 2^n \) density matrix---a positive semi-definite, trace-one operator. An alternative perspective is to express quantum states using \emph{quasi-probability distributions} over a classical-like space. This is achieved by expanding the density matrix in terms of an \emph{operator frame} \cite{renes2004symmetric, Wootters1987, Gross2006}. A \emph{frame} is a generalization of a basis: a spanning set that may be overcomplete but still allows for a faithful reconstruction of any operator. In this context, a frame is a set of Hermitian operators \( \{ F_\lambda \} \), indexed by \( \lambda \in \Lambda \), such that any density matrix \( \rho \) admits an expansion
\begin{equation}
\rho = \sum_{\lambda \in \Lambda} \mu(\lambda) F_\lambda,
\end{equation}
where the coefficients \( \mu(\lambda) \in {\mathbf{R}} \) encode the state. The frame \( \{ F_\lambda \} \) is typically chosen so that the map
\begin{equation}
\rho \mapsto \mu(\lambda) := \operatorname{Tr}[\rho G_\lambda]
\end{equation}
is linear and invertible, with \( \{ G_\lambda \} \) the \emph{dual frame} satisfying
\begin{equation}
\sum_{\lambda \in \Lambda} \operatorname{Tr}[F_\lambda G_{\lambda'}] F_\lambda = F_{\lambda'}.
\end{equation}
The coefficients $\mu(\lambda)$ define a \emph{quasi-probability distribution} over the index set $\Lambda$, meaning a real-valued function that sums to one but is not required to be nonnegative, in contrast to classical probability distributions. When \( \mu(\lambda) \geq 0 \) for all physical states \( \rho \), we obtain a \emph{positive frame representation}. In this representation, quantum channels are always realized as affine maps on the frame coefficients, implemented via \emph{quasi-stochastic matrices}—real matrices whose columns sum to one, but which may contain negative entries. This is the price we always pay if we want to express quantum theory as a stochastic theory \cite{FerrieEmerson2008,FerrieEmerson2009, Spekkens2008, Ferrie2011}.

A symmetric informationally complete (SIC) frame for a qubit \cite{Renes2004} consists of four rank-one operators \( \{ \Pi_\lambda \}_{\lambda=1}^4 \) such that $\Pi_{\lambda}=\frac{1}{4}(1+\mathbf{n}_{\lambda}\cdot\mathbf{\sigma})$, where $\mathbf{\sigma}$ is a vector of Pauli matrices and  $\mathbf{n}_{\lambda}$ form a three dimensional tetrahedron. These satisfy:
\begin{equation}
\operatorname{Tr}[\Pi_\lambda \Pi_{\lambda'}] = 
\begin{cases}
\frac{1}{4} & \text{if } \lambda = \lambda', \\
\frac{1}{12} & \text{if } \lambda \ne \lambda'.
\end{cases}
\end{equation}
A qubit density matrix \( \rho \) can be represented by four real numbers \( p_\lambda \) defined by
\begin{equation}
p_\lambda := \operatorname{Tr}[\rho \, \Pi_\lambda], \qquad \text{with} \quad \sum_\lambda p_\lambda = 1.
\end{equation}
Each \( p_\lambda \) is a genuine probability: \( p_\lambda \geq 0 \) for all physical states \( \rho \). The map \( \rho \mapsto \{ p_\lambda \} \) is linear and injective, providing a positive representation of quantum states in terms of classical probability distributions. The state can be reconstructed via:
\begin{equation}
\rho = \sum_{\lambda=1}^4 \left( 4 p_\lambda - 1 \right) \Pi_\lambda.
\end{equation}
Under a unitary transformation \( U \), the SIC-represented state transforms as:
\begin{equation}
\tilde{p}_{\lambda} = \sum_{\lambda'} T(\lambda \mid \lambda') \, p_{\lambda'},
\end{equation}
where
\begin{equation}
T(\lambda \mid \lambda') := \operatorname{Tr}[U \Pi_{\lambda'} U^\dagger \, \Pi_\lambda]
\end{equation}
is a quasi-stochastic matrix. Although the entries of \( T \) may be negative, the matrix is \emph{column-normalized}, meaning that for each fixed \( \lambda' \), the sum \( \sum_\lambda T(\lambda \mid \lambda') = 1 \). As a result, any normalized input distribution \( \{ p_{\lambda'} \} \) is mapped to a non-negative, normalized output distribution \( \{ \tilde{p}_\lambda \} \), with \( \sum_\lambda p_\lambda' = 1 \). Thus, in the SIC representation: (1) Quantum states remain positive and normalized: \( p_\lambda \geq 0, \sum p_\lambda = 1 \). (2) Quantum channels (e.g., unitaries) become linear maps, possibly involving negativity. (3) Quantum nonclassicality corresponding to negativity shifts from states to transformations. This construction easily generalizes to \( n \)-qubit systems \cite{Us1,Us2}.

To connect this SIC representation more explicitly with a classical-like probability space, we now relabel each index \( \lambda \in \{1,2,3,4\} \) as a pair of bits \( (a, a') \in \{ \pm 1 \}^2 \). Each such pair is associated with a normalized 3D vector \( \mathbf{n}_{aa'} \in {\mathbf{R}}^3 \) pointing toward one of the four vertices of a regular tetrahedron inscribed in the Bloch sphere. These vectors satisfy:
\begin{equation}
\mathbf{n}_{aa'} \cdot \mathbf{n}_{a''a'''} =
\begin{cases}
1 & \text{if } (a, a') = (a'', a'''), \\
- \frac{1}{3} & \text{otherwise}.
\end{cases}
\end{equation}
For convenience they can be chosen as $\mathbf{n}_{aa'}=\frac{1}{\sqrt{3}}(a,a',aa')$, a convention we adopt in this paper.

Let \( \mathbf{s} \in \mathbf{R}^3 \) be the usual Bloch vector associated with a qubit state \( \rho \), so that
\begin{equation}
\rho = \frac{1}{2} \left( \mathbf{I} + \mathbf{s} \cdot \boldsymbol{\sigma} \right),
\end{equation}
where \( \boldsymbol{\sigma} = (\sigma_x, \sigma_y, \sigma_z) \) are the Pauli matrices. Then the SIC probabilities can be written as:
\begin{equation}
p(aa') := \operatorname{Tr}[\rho \, \Pi_{aa'}] = \frac{1}{4} \left( 1 + \mathbf{s} \cdot \mathbf{n}_{aa'} \right),
\end{equation}
where each \( \Pi_{aa'} \) is the SIC projector associated with the pair \( (a, a') \). This gives a linear map from the Bloch vector to a classical-looking probability distribution over four outcomes:
\begin{equation}
\sum_{aa'} p(aa') = 1, \qquad p(aa') \geq 0 \text{ for all physical } \mathbf{s} \text{ such that } |\mathbf{s}| \leq 1.
\end{equation} The state can be reconstructed from the SIC probabilities via:
\begin{equation}
\mathbf{s} = 3 \sum_{aa'} p(aa') \, \mathbf{n}_{aa'},
\end{equation}
which leads back to
\begin{equation}
\rho = \frac{1}{2} \left( \mathbf{I} + 3 \sum_{aa'} p(aa') \, \mathbf{n}_{aa'} \cdot \boldsymbol{\sigma} \right).
\end{equation}

In this representation: (1) The space of qubit states is mapped into the simplex spanned by the four vectors \( \mathbf{n}_{aa'} \). (2) The mapping \( \rho \mapsto p(aa') \) is linear and positivity-preserving. (3) All density matrices correspond to positive probability distributions over \( (a, a') \). Thus, the SIC representation of a qubit state not only admits a fully classical-looking probability interpretation, but also embeds the quantum Bloch ball inside a classical simplex, with no need for negative probabilities. The nonclassicality of the theory emerges only when we consider how these probabilities transform under quantum operations.


A unitary operator \( U \in \mathrm{SU}(2) \) acts on the density matrix by conjugation: \( \rho \mapsto U \rho U^\dagger \). At the level of the Bloch vector, this induces a rotation:
\begin{equation}
\mathbf{s} \mapsto \mathbf{s}' = O \mathbf{s},
\end{equation}
where \( O \in \mathrm{SO}(3) \) is the corresponding rotation matrix. Under this transformation, the SIC probabilities evolve as:
\begin{equation}
p'(bb') = \frac{1}{4} \left( 1 + \mathbf{s}' \cdot \mathbf{n}_{bb'} \right) = \frac{1}{4} \left( 1 + \mathbf{s} \cdot O^T \mathbf{n}_{bb'} \right),
\end{equation} where $T$ denotes matrix transposition. On the other hand, from the perspective of the probability vector \( p(aa') \), the transformation must be linear:
\begin{equation}
p'(bb') = \sum_{aa'} T(bb' \mid aa') \, p(aa').
\end{equation} Substituting the expression for \( p(aa') \), we obtain:
\begin{equation}
\sum_{aa'} T(bb' \mid aa') \cdot \left( \frac{1}{4} (1 + \mathbf{s} \cdot \mathbf{n}_{aa'}) \right) = \frac{1}{4} \left( 1 + \mathbf{s} \cdot O^T \mathbf{n}_{bb'} \right).
\end{equation} This identity must hold for all Bloch vectors \( \mathbf{s} \), so by comparing terms, we find:
\begin{eqnarray}
\sum_{aa'} T(bb' \mid aa') &= \frac{1}{4}, \nonumber\\
\sum_{aa'} T(bb' \mid aa') \, \mathbf{n}_{aa'} &= \frac{1}{4} \, O^T \mathbf{n}_{bb'}.
\end{eqnarray} The above equations fully determine the transition matrix \( T(bb' \mid aa') \). Explicitly, we can write:
\begin{equation}
T(bb' \mid aa') = \frac{1}{4} \left( 1 + 3 \, \mathbf{n}_{aa'} \cdot O^T \mathbf{n}_{bb'} \right).
\end{equation} Although the matrix \( T \) may contain negative entries, it preserves both normalization and positivity when acting on valid quantum states. That is, for any Bloch vector \( \mathbf{s} \) with \( |\mathbf{s}| \leq 1 \), the resulting distribution
\begin{equation}
p'(bb') = \sum_{aa'} T(bb' \mid aa') \, p(aa') = \frac{1}{4} \left( 1 + \mathbf{s}' \cdot \mathbf{n}_{bb'} \right)
\end{equation}
remains nonnegative and sums to one. In this sense, the transformation is \emph{quasi-stochastic}: each column of \( T \) sums to one,
\begin{equation}
\sum_{bb'} T(bb' \mid aa') = 1,
\end{equation}
but some individual entries may be negative. The key point is that, despite this negativity, the map is positivity-preserving on the convex set of physical qubit states but not on the full probabilistic simplex (in particular not on the deterministic probability distributions such as $p(aa')=\frac{1}{4}(1+a)(1+a')$). Negativity arises not in the states, but in the structure of the transformation itself.

To see this explicitly, consider the inner product between two tetrahedral vectors. Since \( \mathbf{n}_{aa'} \cdot \mathbf{n}_{bb'} = -1/3 \) for distinct pairs, it follows that unless \( O \) maps each \( \mathbf{n}_{aa'} \) exactly onto another tetrahedral vector (i.e., a permutation of the frame), some inner products \( \mathbf{n}_{aa'} \cdot O^T \mathbf{n}_{bb'} \) will fall below \( -1/3 \), making \( T(bb' \mid aa') \) negative. More generally, any quantum channel -- that is, any completely positive, trace-preserving (CPTP) map represented by a Kraus decomposition -- acts affinely on the Bloch vector. In the SIC representation, this means the output probabilities \( p'(aa') \) are still of the form
\begin{equation}
p'(aa') = \frac{1}{4} \left( 1 + \mathbf{s}' \cdot \mathbf{n}_{aa'} \right),
\end{equation}
where the transformed Bloch vector \( \mathbf{s}' \) is given by
\begin{equation}
\mathbf{s}' = A \mathbf{s} + \mathbf{t},
\end{equation}
for some real \( 3 \times 3 \) matrix \( A \) and translation vector \( \mathbf{t} \in \mathbf{R}^3 \). This affine structure follows directly from the linearity and trace-preservation of quantum operations. The transformation \( A \mathbf{s} + \mathbf{t} \) maps the Bloch ball into itself, possibly contracting or shifting it, but never taking valid states outside the set of positive distributions in the SIC frame.

Therefore: (1) The transformation matrix \( T \) is quasi-stochastic: real, column-normalized, but not necessarily nonnegative. (2) The only unitaries that yield a genuinely stochastic (positive) transition matrix are those that permute the SIC projectors up to a global phase. (3) In general, \emph{negativity is inevitable in the dynamics}, even though the quantum states are represented as positive probability distributions.

We would like to point out a striking feature of the SIC representation: even entangled quantum states are described by semi-positive probability distributions. This may appear counterintuitive, as entanglement is widely regarded as a marker of quantum nonclassicality. Nevertheless, it follows directly from the geometry of the SIC frame. Consider, for example, the maximally entangled singlet state:
\begin{equation}
|\psi_-\rangle = \frac{1}{\sqrt{2}} (|0\rangle \otimes |1\rangle - |1\rangle \otimes |0\rangle).
\end{equation}
In the local SIC representation, this state corresponds to the joint probability distribution
\begin{equation}
p(aa', bb') = \frac{1}{16} \left( 1 - \mathbf{n}_{aa'} \cdot \mathbf{n}_{bb'} \right),
\end{equation}
where \( \mathbf{n}_{aa'} \) and \( \mathbf{n}_{bb'} \) are local SIC vectors on Alice's and Bob's sides, respectively. Since the dot product of any two tetrahedral unit vectors satisfies \( |\mathbf{n}_{aa'} \cdot \mathbf{n}_{bb'}| \leq 1 \), the expression is non-negative for all $(aa',bb')$. This shows that entanglement, even in its maximally nonlocal form, does not require negativity in the SIC distribution. However, this positivity does not imply classical behavior. To test Bell inequalities such as CHSH, one must evaluate correlations under different measurement settings, which correspond to applying different local rotations to the SIC frame. These rotations are implemented by quasi-stochastic transformations. Even though the underlying state \( p(aa', bb') \) remains non-negative, the transformed distributions -- corresponding to each setting combination -- yield quantum correlations that can violate the CHSH bound. The violation arises not from the negativity of the state, but from the structure of the measurement transformations applied to it. In this way, quantum nonclassicality re-emerges at the level of process, not state. 

\section{Generalised Bloch Vector}

Equipped with the frame formalism from the previous section, we now introduce the concept of a \emph{generalized Bloch vector} (GBV, for short). It will be crucial to understand the main result of this paper.

Consider an arbitrary non-negative probability distribution over \( n \) binary variables \( a_i = \pm 1 \), $p(a_1, a_2, \dots, a_n),$ where \( i = 1, 2, \dots, n \).

Such a distribution is fully characterized by its correlation functions (or moments), including the single-variable averages \( \langle a_i \rangle \), pairwise correlations \( \langle a_i a_j \rangle \), and higher-order correlators up to the full \( n \)-point correlation \( \langle a_1 a_2 \dots a_n \rangle \). It can be expressed as:
\begin{equation}
p(a_1, a_2, \dots, a_n) = \frac{1}{2^n} \left( 1 + \sum_{i=1}^n \langle a_i \rangle a_i + \sum_{i < j} \langle a_i a_j \rangle a_i a_j + \dots + \langle a_1 a_2 \dots a_n \rangle a_1 a_2 \dots a_n \right).
\end{equation}

We introduce a set of $2^n$ normalized vectors \( \mathbf{n}_{a_1a_2\dots a_n} \in \mathbf{R}^{2^n - 1} \) associated with each $n$-bit string \( (a_1, a_2, \dots, a_n) \), defined by:
\begin{equation}
\mathbf{n}_{a_1a_2\dots a_n} = \frac{1}{\sqrt{2^n - 1}} \left( a_1, a_2, \dots, a_n, a_1 a_2, a_1 a_3, \dots, a_1 a_2 \dots a_n \right),
\end{equation}
where the components include all nontrivial monomials (excluding the constant term \( 1 \)) formed by products of the bits. The dimension of this space is \( 2^n - 1 \). At this stage, the arrangement of monomials is immaterial; any order may be chosen. The scalar product between two such vectors is given by:
\begin{equation}
\mathbf{n}_{a_1a_2\dots a_n} \cdot \mathbf{n}_{a_1'a_2'\dots a_n'} = 
\begin{cases}
1 & \text{if } (a_1, a_2, \dots, a_n) = (a_1', a_2', \dots, a_n'), \\
-\frac{1}{2^n - 1} & \text{otherwise}.
\end{cases}
\end{equation} These vectors form an overcomplete frame, as seen from the identity:
\begin{equation}
\sum_{a_1, a_2, \dots, a_n = \pm 1} \mathbf{n}_{a_1a_2\dots a_n}^T \, \mathbf{n}_{a_1a_2\dots a_n} = \frac{2^n}{2^n - 1} \, \mathbf{I},
\end{equation}
where $\mathbf{n}_{a_1a_2\dots a_n}^T \, \mathbf{n}_{a_1a_2\dots a_n}$ is a dyadic product of two vectors and \( \mathbf{I} \) denotes the \( (2^n - 1) \times (2^n - 1) \) identity matrix. This leads to a compact representation of any such probability distribution in terms of a generalized Bloch vector \( \mathbf{w} \in \mathbf{R}^{2^n - 1} \):
\begin{equation}
p(a_1, a_2, \dots, a_n) = \frac{1}{2^n} \left( 1 + \mathbf{w} \cdot \mathbf{n}_{a_1a_2\dots a_n} \right),
\label{generalp}
\end{equation}
where \( \mathbf{w} \) contains all the nontrivial correlators of the distribution and plays the role of a classical analogue of the Bloch vector. Note that \( \mathbf{w} \) lies within the probabilistic simplex spanned by convex combinations of the vectors \( (2^n - 1)\mathbf{n}_{a_1a_2\dots a_n} \). Each such vector corresponds to a deterministic probability distribution -- that is, one with zero entropy.

Given a known probability distribution \( p(a_1, a_2, \dots, a_n) \), the generalized Bloch vector \( \mathbf{w} \in \mathbf{R}^{2^n - 1} \) can be reconstructed using the overcompleteness relation of the frame vectors \( \mathbf{n}_{a_1a_2\dots a_n} \). From the decomposition in (\ref{generalp})
we isolate the frame component:
\begin{equation}
\mathbf{w} \cdot \mathbf{n}_{a_1a_2\dots a_n} = 2^n p(a_1, a_2, \dots, a_n) - 1.
\end{equation} To recover \( \mathbf{w} \), we multiply both sides by \( \mathbf{n}_{a_1a_2\dots a_n} \) and sum over all configurations, using the frame identity:
\begin{equation}
\sum_{a_1, \dots, a_n = \pm 1} \mathbf{n}_{a_1a_2\dots a_n}^T \mathbf{n}_{a_1a_2\dots a_n} = \frac{2^n}{2^n - 1} \, \mathbf{I}.
\end{equation} This yields:
\begin{equation}
\sum_{a_1, \dots, a_n} \left( 2^n p(a_1, \dots, a_n) - 1 \right) \mathbf{n}_{a_1a_2\dots a_n} = \frac{2^n}{2^n - 1} \, \mathbf{w},
\end{equation}
and therefore:
\begin{equation}
\mathbf{w} = \frac{2^n - 1}{2^n} \sum_{a_1, \dots, a_n = \pm 1} \left( 2^n p(a_1, \dots, a_n) - 1 \right) \mathbf{n}_{a_1a_2\dots a_n}.
\end{equation}
As the sum of all vectors $\mathbf{n}_{a_1a_2\dots a_n}$ vanishes, the $-1$ is retained in the formula purely for clarity.

Now consider a general stochastic process \( T \) acting on probability distributions. Such a process maps \( p \mapsto \tilde{p} \), i.e.,
\begin{equation}
\tilde{p}(b_1, \dots, b_n) = \sum_{a_1, \dots, a_n} T(b_1, \dots, b_n \mid a_1, \dots, a_n) \, p(a_1, \dots, a_n),
\end{equation}
where \( T \) is a conditional probability distribution: for each fixed \( a = (a_1, \dots, a_n) \), \( \sum_b T(b \mid a) = 1 \), and \( T(b \mid a) \geq 0 \). Substituting the Bloch expansion of \( p(a) \) into the equation for \( \tilde{p}(b) \), we obtain:
\begin{align}
\tilde{p}(b) &= \sum_a T(b \mid a) \left( \frac{1}{2^n} \left( 1 + \mathbf{w} \cdot \mathbf{n}_a \right) \right) \\
&= \frac{1}{2^n} \sum_a T(b \mid a) + \frac{1}{2^n} \sum_a T(b \mid a) \, (\mathbf{w} \cdot \mathbf{n}_a).
\end{align} Define:
\begin{equation}
c(b) := \frac{1}{2^n} \sum_a T(b \mid a), \quad \text{and} \quad M_b := \frac{1}{2^n} \sum_a T(b \mid a) \, \mathbf{n}_a,
\end{equation}
so that:
\begin{equation}
\tilde{p}(b) = c(b) + M_b \cdot \mathbf{w}.
\end{equation}
Now, we can again write \( p'(b) \) in the Bloch form:
\begin{equation}
p'(b) = \frac{1}{2^n} \left( 1 + \mathbf{w}' \cdot \mathbf{n}_b \right),
\end{equation}
which implies:
\begin{equation}
\mathbf{w}' \cdot \mathbf{n}_b = 2^n p'(b) - 1 = 2^n \left( c(b) + M_b \cdot \mathbf{w} \right) - 1.
\end{equation}
Since this is true for all \( b \), it follows that \( \mathbf{w}' \) is a linear (affine) function of \( \mathbf{w} \), i.e.,
\begin{equation}
\mathbf{w}' = A \mathbf{w} + \mathbf{t},
\end{equation}
for some real matrix \( A \) and vector \( \mathbf{t} \in \mathbf{R}^{2^n - 1} \).

We see that any stochastic process acting on such probability distributions induces an affine transformation of the generalized Bloch vector \( \mathbf{w} \). This is analogous to the action of quantum channels on the Bloch vector of a qubit, but here realized entirely in a classical probabilistic setting.

The generalized Bloch representation extends naturally to quasi-stochastic processes, where the transition map \( T(b \mid a) \) is normalized but may take negative values:
\begin{equation}
\sum_b T(b \mid a) = 1, \qquad T(b \mid a) \not\geq 0.
\end{equation} As before, the transformation acts linearly on the probability distribution:
\begin{equation}
p'(b) = \sum_a T(b \mid a)\, p(a),
\end{equation}
which in Bloch form becomes:
\begin{equation}
p'(b) = \frac{1}{2^n} \left( 1 + \mathbf{w}' \cdot \mathbf{n}_b \right), \qquad \text{with } \mathbf{w}' = A \mathbf{w} + \mathbf{t}.
\end{equation} Thus, any quasi-stochastic process induces an affine map on the generalized Bloch vector.

\medskip
\noindent

\section{Quasi-Bistochastic Model of Quantum Measurement}

We introduce a minimal model of quantum measurement, in which a qubit is directly coupled to a classical bit acting as a measurement register. The measurement is described by a quasi-stochastic process: it acts linearly on classical, positive probability distributions but may involve negative transition weights, preventing its interpretation as a conventional stochastic map. Crucially, both system and apparatus are represented within a finite classical probability space, avoiding the need for a quantum–classical boundary or a von Neumann-type entanglement chain. While the model reproduces standard quantum measurement statistics and collapse behavior, its internal structure reflects the nonclassical features of quantum theory—not in the state space, but in the measurement dynamics itself.

Our system consists of a qubit represented by a positive probability distribution $p(aa')$ over two classical bits $a,a' =\pm 1$ and a single bit meter \( \alpha = \pm 1\), initialized deterministically:
\begin{equation}
p(\alpha) = \frac{1}{2}(1 + \alpha).
\end{equation} leading to an initial distribution over the three-bit system
\begin{equation}
p(\alpha, a, a') = p(\alpha)p(a, a').
\end{equation} We use the generalized Bloch vector to describe this initial distribution. Since we deal with three bits, we require 7-dimensional simplex vectors
\begin{equation}
\mathbf{n}_{\alpha a a'} = \frac{1}{\sqrt{7}}(a, a', a a', \alpha a, \alpha a', \alpha a a', \alpha).
\end{equation}
Here, we choose a particular ordering of monomials to facilitate intuitive insight into the model. It is also convenient to rewrite these vectors as:
\begin{equation}
\mathbf{n}_{\alpha a a'} = \sqrt{\frac{3}{7}} \left( \mathbf{n}^{(1)}_{a a'} + \alpha \mathbf{n}^{(2)}_{a a'} \right) + \frac{\alpha}{\sqrt{7}} \, \mathbf{e}^{(3)},
\end{equation}
where the vectors \( \mathbf{n}^{(1)}_{a a'} \), \( \mathbf{n}^{(2)}_{a a'} \), and \( \mathbf{e}^{(3)} \) lie in three mutually orthogonal subspaces of dimensions 3, 3, and 1, respectively. These are defined as:
\begin{equation}
\mathbf{n}^{(1)}_{a a'} = \frac{1}{\sqrt{3}}(a, a', a a', 0, 0, 0, 0), \quad
\mathbf{n}^{(2)}_{a a'} = \frac{1}{\sqrt{3}}(0, 0, 0, a, a', a a', 0), \quad
\mathbf{e}^{(3)} = (0, 0, 0, 0, 0, 0, 1).
\end{equation}
Thus, our initial distribution has the following generalized Bloch vector $\mathbf{w}=\sqrt{\frac{7}{3}}(\mathbf{s}^{(1)}+\mathbf{s}^{(2)})+\sqrt{7}\mathbf{e}^{(3)}$, where $\mathbf{s}^{(i)}$ are the same qubit Bloch vectors but in different, orthogonal subspaces. Note that if the classical bit is initialized with the probability distribution $p(\alpha)=\frac{1}{2}(1+\langle\alpha\rangle\alpha)$ we get $\mathbf{w}=\sqrt{\frac{7}{3}}(\mathbf{s}^{(1)}+\langle\alpha\rangle\mathbf{s}^{(2)})+\langle\alpha\rangle\sqrt{7}\mathbf{e}^{(3)}$. However, we are only interested in the classical bit initializations such that $\langle\alpha\rangle=\pm 1$ or else we deal with a noisy meter. This remark is important to understand the next paragraph. We finally have 
\begin{equation}
 p(\alpha, a, a') = p(\alpha)p(a, a')=\frac{1}{8}(1+\mathbf{w}\cdot \mathbf{n}_{\alpha a a'}).   
\end{equation} 

Let us now interpret the three orthogonal subspaces appearing in our representation. The first subspace corresponds to the state/description of the qubit. The second encodes qubit–bit correlations in the sense that we store here the initial preparation of the meter with respect to the qubit - it is either $\langle\alpha\rangle=+1$ or $\langle\alpha\rangle=-1$. The third represents the measurement bit register.

In our model, a von Neumann projective measurement transforms the initial probability distribution, represented by the generalized Bloch vector \( \mathbf{w} \), into a post-measurement distribution in which the qubit collapses to a pure state aligned with the measurement direction \( \mathbf{m} \) on the Bloch sphere. The distribution of measurement outcomes must reproduce the correct quantum statistics: \( p(\beta) = \frac{1}{2} (1 + \beta \, \mathbf{s} \cdot \mathbf{m}) \). This must translate to the post measurement vector $\mathbf{w}'=\sqrt{\frac{7}{3}}((\mathbf{s}\cdot\mathbf{m}) \mathbf{m}^{(1)}+\mathbf{m}^{(2)})+\sqrt{7}(\mathbf{m}\cdot\mathbf{s}) \mathbf{e}^{(3)}$. From the previous section, we know that any quasi-stochastic process acts as an affine transformation -- denoted  $A_{\mathbf{m}}$ -- on the generalized Bloch vector. Since this vector consists of three distinct components, as previously shown, we can now postulate a necessary behaviour of such a transformation. Using formulas derivesd in the previous sections we can see that in our case, the process reads 
\begin{equation}
S[\beta,b,b'|\alpha,a,a']=\frac{1}{8}(1+7 \mathbf{n}_{\beta bb'}\cdot \mathbf{A}_{\mathbf{m}}\mathbf{n}_{\alpha aa'}).
\end{equation}

The most general form of $A_{\mathbf{m}}$ reads

\begin{equation}
   A_{\mathbf{m}} = (\mathbf{m}^{(1)})^T\mathbf{r^{(1)}}(x) + (\mathbf{m}^{(2)})^T\mathbf{r^{(2)}}(y)+(\mathbf{e}^{(3)})^T\mathbf{r^{(3)}}(z),
\end{equation}
where 
\begin{eqnarray}
    \mathbf{r^{(1)}}(x) &=& x \mathbf{m}^{(1)}+(1-x)\mathbf{m}^{(2)}\nonumber\\
     \mathbf{r^{(2)}}(y) &=& y (\mathbf{m}^{(1)}-\mathbf{m}^{(2)})+\frac{1}{\sqrt{3}}\mathbf{e}^{(3)}\nonumber\\
    \mathbf{r^{(3)}}(z) &=& z \mathbf{m}^{(1)}+(\sqrt{3}-z)\mathbf{m}^{(2)}.  
\end{eqnarray}
Here, $x,y,z$ are arbitrary real numbers. They generate a family of quasi-bistochastic processes, giving a proper qubit-meter behaviour if applied once. It can be checked that they do not play a role at this stage.
However, we must make sure that a repeated application of the measurement process does not produce negative probability distributions. In the second iteration of the process, the parameters $x,y,z$ appear and may lead to a negative qubit-meter probability distribution. We need to make a proper choice. A convenient one, as we will see below, is $x=1,y=0,z=0$, giving us

\begin{equation}
A_{\mathbf{m}} = (\mathbf{m}^{(1)})^T \mathbf{m}^{(1)} + \frac{1}{\sqrt{3}} \, (\mathbf{m}^{(2)})^T \mathbf{e}^{(3)} + \sqrt{3} \, (\mathbf{e}^{(3)})^T \mathbf{m}^{(2)}.
\end{equation}
First, we notice that $A_{\mathbf{m}}^{2k}=A_{\mathbf{m}}^{2}$
and $A_{\mathbf{m}}^{2k+1}=A_{\mathbf{m}}$
for $k \geq 1$. This leads to a repetitive behaviour where the initial qubit-meter GBV $\mathbf{w}$ transforms to $\mathbf{w}'=\sqrt{\frac{7}{3}}((\mathbf{s}\cdot\mathbf{m}) \mathbf{m}^{(1)}+\mathbf{m}^{(2)})+\sqrt{7}(\mathbf{m}\cdot\mathbf{s}) \mathbf{e}^{(3)}$ and if applied once more it becomes $\mathbf{w}''=(\mathbf{s}\cdot\mathbf{m}) \sqrt{\frac{7}{3}}(\mathbf{m}^{(1)}+\mathbf{m}^{(2)})+\sqrt{7} \mathbf{e}^{(3)}$. The third application throws $\mathbf{w}''$ back to $\mathbf{w}'$ etc. Both $\mathbf{w}'$ and $\mathbf{w}''$ generate positive probability distributions. 

It is instructive to re-write the above behaviour directly in the language of quasi-bistochastic process generated by $A_{\mathbf{m}}$. It reads 

\begin{equation}
S[\beta bb'|\alpha aa']= \frac{1}{8}(1+
\alpha\beta\mathbf{m}\cdot\mathbf{n}_{bb'})(1+3\alpha\beta\mathbf{m}\cdot\mathbf{n}_{aa'}).
\label{theprocess}
\end{equation}
This process transforms the initial qubit-meter distribution $p(aa')\frac{1}{2}(1+\alpha)$ to 
\begin{equation}
\sum_{\alpha,a,a'}S[\beta bb'|\alpha aa']p(aa')\frac{1}{2}(1+\alpha) = \frac{1}{8}(1+\beta \mathbf{m}\cdot\mathbf{n}_{bb'})(1+\beta\mathbf{m}\cdot\mathbf{s}),
\end{equation}
where $\mathbf{s}$ is the qubit's Bloch vector. Second application of the process gives $\frac{1}{8}(1+(\mathbf{m}\cdot\mathbf{s}) (\mathbf{m}\cdot\mathbf{n}_{bb'}))(1+\beta)$. It can be easily seen these are positive probability distributions. Positivity of the qubit-meter probability is also maintained for an arbitrary chain of successive measurements with different directions, i.e., $\Pi_{i=1}^NA_{\mathbf{m_i}}$. The process (\ref{theprocess}) has negative transition probabilities for any measurement direction $\mathbf{m}$ as expected. The negativity arises solely from the term $1+3\alpha\beta\mathbf{m}\cdot\mathbf{n}_{aa'}$ since there exist $a,a'$ such that $|\mathbf{m}\cdot\mathbf{n}_{aa'}|\geq \frac{1}{3}$ for all unit vectors $\mathbf{m}$. 

For an arbitray set of $x,y,z$ we have a family of quasi-bistochastic processes 
\begin{eqnarray}
&&S_{xyz}[\beta bb'|\alpha aa'] =\nonumber\\
&&\frac{1}{8}(1+\sqrt{3}\beta(z+(\sqrt{3}-z)\alpha)\mathbf{m}\cdot\mathbf{n}_{aa'}+\alpha\beta \mathbf{m}\cdot\mathbf{n}_{bb'}+\nonumber\\
&&3(x+(1-x)\alpha+\beta y(1-\alpha))\mathbf{m}\cdot\mathbf{n}_{bb'}\mathbf{m}\cdot\mathbf{n}_{aa'}). 
\label{gprocess}
\end{eqnarray}  
Since every one of them leads to a proper collapse behaviour if applied once, there is no range of $x,y,z$ giving us a bistochastic process, i.e., a process with strictly non-negative transition probabilities. If this happened we would have a classical simulation of a von Neumann measurement.

This family can be written in the following way 

\begin{eqnarray}
&&S_{xyz}[\beta bb'|\alpha aa']=\nonumber\\
&&\frac{1}{8}\Big[
\big(1+\sqrt{3}\,\beta\big(z+(\sqrt{3}-z)\alpha\big)\,
\mathbf{m}\!\cdot\!\mathbf{n}_{aa'}\big)
\big(1+\alpha\beta\,\mathbf{m}\!\cdot\!\mathbf{n}_{bb'}\big)
+\nonumber\\
&&\Big(
3\big(x+(1-x)\alpha+\beta\,y(1-\alpha)\big)
-\sqrt{3}\,\alpha\big(z+(\sqrt{3}-z)\alpha\big)
\Big)
\big(\mathbf{m}\!\cdot\!\mathbf{n}_{aa'}\big)
\big(\mathbf{m}\!\cdot\!\mathbf{n}_{bb'}\big)
\Big].
\end{eqnarray}
We get (\ref{theprocess}) if and only if the term $\Big(
3\big(x+(1-x)\alpha+\beta\,y(1-\alpha)\big)
-\sqrt{3}\,\alpha\big(z+(\sqrt{3}-z)\alpha\big)
\Big)
\big(\mathbf{m}\!\cdot\!\mathbf{n}_{aa'}\big)
\big(\mathbf{m}\!\cdot\!\mathbf{n}_{bb'}\big)
\Big]$ vanishes, which can only happen for $x=1,y=z=0$. This uniqueness reflects the semi-cyclic nature of the matrix $A_{\mathbf{m}}$ discussed earlier.
It parallels the behaviour of the CNOT gate in the von~Neumann measurement model: the measured qubit couples to the measurement register through a controlled operation that, when applied twice, becomes the identity. In our formalism, this symmetry manifests as a two-step alternation between the pre- and post-measurement forms of the generalized Bloch vector.

We leave it as an open problem if there are other members of the family that preserve the positivity of the qubit-meter distribution when the measurement is repeated and arbitrary number of times.

It is instructive to examine the collision entropy of the qubit-bit distribution before and after the measurement process \cite{Us3}. We choose collision entropy because it is naturally related to the purity of probability distributions, i.e., how close a given probability distribution is to a distribution with one certain outcome (maximal purity). Collision entropy is Renyi entropy 
\begin{equation}
H_\alpha(p) = \frac{1}{1 - \alpha} \log_{2} \left( \sum_i p_i^\alpha \right).
\end{equation} with $\alpha=2$. For a distribution represented by a generalized Bloch vector \( \mathbf{w} \), the collision entropy takes the form:
\begin{equation}
H_{2} = -\log_{2}\left(\frac{1}{8} \left( 1 + \frac{1}{7}|\mathbf{w}|^2 \right)\right).
\end{equation}
The difference between the after measurement entropy and the initial entropy reads 
\begin{equation}
H_2^{after}-H_2^{before} = -1+\log_2{\frac{3+|\mathbf{s}|^2}{1+|\mathbf{m}\cdot\mathbf{s}|}}.
\end{equation}
We can see that if the measurement direction is aligned with the direction of Bloch vector $\mathbf{s}$ the difference is minimal and if the state is pure it becomes zero. On the other hand, if the measurement direction is orthogonal to the Bloch vector representing a pure state we get the entropy increase of exactly one bit. 

\section{Comparison to classical measurement}

We need to ask a question: what makes this quantum measurement model quantum?
The telltale sign is that it is described by a quasi-bistochastic process. But which part of the model is responsible for this feature? Where exactly does quasi-stochasticity enter? 

The process discussed here takes the form $S[\beta,b,b'|\alpha,a,a']=\frac{1}{8}(1+7 \mathbf{n}_{\beta bb'}\cdot \mathbf{A}_{\mathbf{m}}\mathbf{n}_{\alpha aa'})$ and thus this is a well defined mathematical problem: we require that for all measurement directions $\mathbf{m}$, $\mathbf{n}_{\beta bb'}\cdot A_{\mathbf{m}}\mathbf{n}_{\alpha aa'}\geq -\frac{1}{7}$. We can immediately see that the necessary condition is that the length of $\mathbf{m}$ is not greater than $\frac{1}{3}$, which is not possible as the vector $\mathbf{m}$ must be normalised by definition. Thus, the measurement direction is the culprit here and we can understand it in the following way.

Since we use only one bit to store the results of a measurement on a two-bit probability distribution, some information loss is to be expected. It is impossible to retrieve the full information about  $p(aa')$  from the post-measurement meter bit’s probability distribution  $\tilde{p}(\alpha)$. The distribution $p(aa')$ contains information about the average value $\langle A\rangle$ of bit $a$ (proportional to the $s_x$ component of the Bloch vector $\mathbf{s}$), $\langle A'\rangle$ of bit $a'$ (proportional to $s_y$) and their correlations $\langle AA'\rangle$ (proportional to $s_z$). Quantum measurement outputs a linear combination of these three averages with weights coming from the vector $\mathbf{m}$, i.e., $s_xm_x+s_ym_y+s_zm_x$. Now, it is easy to find a positive stochastic process that outputs convex combinations of $s_x,s_y$ and $s_z$. Such a process is a probabilistic mixture of positive stochastic processes: (1) $p(aa')p(\alpha)\rightarrow p(aa')p(a\alpha)$ (2) $p(aa')p(\alpha)\rightarrow p(aa')p(a'\alpha)$ and (3) $p(aa')p(\alpha)\rightarrow p(aa')p(aa'\alpha)$ such that the final process reads $p(aa')p(\alpha)\rightarrow p(aa')(r_1p(a\alpha)+r_2p(a'\alpha)+r_3p(aa'\alpha))$, where each $r_i$ is non-negative and  $r_1+r_2+r_3=1$. However we cannot have a non-trivial, normalized measurement vector whose components $m_x,m_y,m_z$ are probabilities since it is the sum of their squares $m_x^2+m_y^2+m_z^2$ that equals one, not their sum with the exception of $\mathbf{m}$ aligned with the vectors $(1,0,0),(0,1,0)$ and $(0,0,1)$.

Suppose, we use two measurement bits $\alpha$ and $\alpha'$ instead. In this case we can find a bistochastic positive operation $p(aa')\frac{1}{2}(1+\alpha)\frac{1}{2}(1+\alpha')\rightarrow p(aa')\frac{1}{2}(1+a\alpha)\frac{1}{2}(1+a'\alpha')$ that simply copies Bloch vector to the meter without a quantum collapse. This way we avoid a problematic measurement direction and we do not need to use quasi-stochastic processes to achieve this. We simply read out Bloch vector characterizing a qubit we observe. To conclude we see that the "quantumness" of projective measurement in our model has it roots in imposing 3D measurement directions in the measurement process.

\section{Discussion and conclusions}
To understand the physical meaning of the process \( S \) introduced in the previous section, let us first 
consider the standard von Neumann measurement in the SIC frame representation. A projective measurement 
along a direction \( \mathbf{m} \) transforms an initial Bloch vector \( \mathbf{s} \) to its projection 
\( (\mathbf{s} \cdot \mathbf{m}) \, \mathbf{m} \). In the SIC picture, this corresponds to the channel:
\begin{equation}
V[bb' \mid aa'] = \frac{1}{4} \left( 1 + 3 \, \mathbf{n}_{bb'} \cdot R \, \mathbf{n}_{aa'} \right),
\end{equation}
where \( R = \mathbf{m}^T \mathbf{m} \).

This is precisely the transformation that results from our model when the measurement bit is discarded:
\begin{equation}
\sum_{\beta, \alpha, a, a'} S[\beta, b, b' \mid \alpha, a, a'] \, p(a, a') \, p(\alpha)
= \sum_{a, a'} V[bb' \mid aa'] \, p(a, a').
\end{equation}
Since \( R \) is rank one, the transformation \( V \) is non-invertible. This reflects a fundamental feature 
of projective measurement: the process eliminates all components orthogonal to \( \mathbf{m} \), and the 
original state cannot be recovered.

A similar situation holds in our qubit--bit coupling model. The transformation \( A_{\mathbf{m}} \) is 
likewise non-invertible, and therefore the full process \( S \) is irreversible. Once the measurement occurs, 
the system’s prior state is lost.

An intriguing feature of our measurement scenario is that it cannot be reproduced within standard quantum 
theory. The reason lies in the structure of the pre-measurement state, which we assume to be a product 
distribution \( p(aa')p(\alpha) \), with \( p(\alpha) = \frac{1}{2}(1 + \alpha) \) representing a deterministic 
classical bit. In quantum theory, even the minimal system---a single qubit---is represented in the SIC 
frame as a probability distribution over two classical bits, say \( (\alpha, \alpha') \). One might attempt to 
prepare a state of the form \( p(\alpha, \alpha') = p(\alpha)\,p(\alpha') \), with the marginal corresponding to 
$\alpha$ a pure distribution $p(\alpha)=\frac{1}{2}(1+\alpha)$ of the meter. However, this is not physically 
realizable, since for any qubit state $\langle\alpha\rangle = \frac{s_x}{\sqrt{3}}$ and $|s_x|\leq 1$. Thus the 
quasi-stochastic nature of the process \( S \) is not itself problematic---such negativity is a well-known and 
intrinsic feature of quantum dynamics. Rather, the issue lies in the initial state: our model allows product 
distributions that quantum theory forbids, even for a single qubit.

Another interesting feature of our measurement process is that it can output proper quantum mechanical 
measurement results to an arbitrary number of observers, preserving the intended collapse. To see this, 
prepare $M$ classical bits distributed as $\Pi_{i=1}^M r(\gamma_i)$, where $r(\gamma_i)=\frac{1}{2}(1+\gamma_i)$. 
Now, our initial state reads $p(aa')p(\alpha)\Pi_{i=1}^M r(\gamma_i)$. Take the following quasi-stochastic process 
$\Pi_{i=1}^M R(\gamma_i|\beta)S(\beta bb'|\alpha aa')$, where 
$R(\gamma_i|\beta)p(\beta\dots)r(\gamma_i)=p(\beta\dots)r(\beta\gamma_i)$.

This observation points to a broader interpretation. The process \( S \) does not merely simulate quantum 
measurement; it offers a minimal, operational link between quantum and classical domains. A single classical 
bit registers the outcome, without invoking decoherence, entanglement, or any macro–micro boundary. In 
this sense, our model acts as a structured interface: it captures the emergence of classical outcomes from 
quantum systems entirely within a discrete, probabilistic framework. Additionally, we see that our minimalist 
construction echoes some of the core ideas behind \emph{Quantum Darwinism} \cite{zurek2009quantum}, where 
classical objectivity arises from the redundant imprinting of quantum information onto the environment, 
allowing multiple observers to access consistent outcomes. Our model achieves a similar end---the production 
of definite classical outcomes---but without environmental redundancy. Instead, classicality is enforced 
directly through a quasi-stochastic process that couples a finite quantum system to an arbitrary number of 
classical registers.

Ultimately, the process \( S \) can be viewed as a compact operational model of the quantum--classical 
boundary: not as a metaphysical divide, but as a structured translation between two probabilistic 
representations of physical reality. The use of negative transition weights, or quasi-probabilities, is not 
foreign to quantum theory—it appears in phase-space formulations, weak measurement frameworks, and 
contextuality proofs. In the present model, such negativity can be traced to the combination of 
three-dimensional measurement directions with the use of a single classical bit as the measurement register, 
rather than to any nonclassicality of the state representation. What this work contributes is a consistent 
extension of that structure across the boundary, treating the classical measurement register within the same 
quasi-stochastic framework that applies to the quantum system. This perspective enables a classically 
structured, one-shot description of measurement that bypasses the infinite regress of the von Neumann 
chain. Several aspects of the construction merit further investigation, including the behaviour of repeated 
measurements for general choices of transformation parameters and the possible characterisation of parameter 
regimes that reduce, though not necessarily minimise, the degree of negativity compatible with projective 
statistics. Whether this framework yields deeper insights into the foundations of quantum theory, or offers 
practical advantages for quantum simulation and control, remains a compelling direction for further 
exploration.

\printbibliography
\end{document}